\documentclass[11pt]{article}
\usepackage{hyperref}
\usepackage{amsmath}
\usepackage{graphicx}% Include figure files
\def\bea{\begin{eqnarray}}
\def\eea{\end{eqnarray}}
\def\beq{\begin{equation}}
\def\eeq{\end{equation}}
\def\er{\eqref}

\def\a{\alpha}
\def\b{\beta}

\def\s{\sigma}

\def\g{\gamma}
\def\G{\Gamma}

\def\d{\delta}
\def\p{\partial}

\def\nab{\nabla}
\def\del{\partial}

\begin{document}

\begin{center} {\Large \bf
Initial value constraints with tensor matter}
\end{center}

\vskip 5mm
\begin{center} \large
%\author
{{Ted Jacobson%\footnote{E-mail: jacobson@umd.edu}
}}
\end{center}

\vskip  0.5 cm
{\centerline{\it Maryland Center for Fundamental Physics}}
{\centerline{\it Department of Physics, University of Maryland}}
%{\centerline{\it University of Maryland}}
{\centerline{\it College Park, MD 20742-4111, USA}}

\vskip 1cm

\begin{abstract}
In generally covariant metric gravity theories with tensor matter
fields, the initial value constraint equations, unlike in general relativity, are 
in general not
just the $0\mu$ components of the metric field equation. This 
happens because higher 
derivatives can occur in the matter stress tensor.  A universal 
form for these constraints is derived here from a generalized
Bianchi identity that includes matter fields. As an application, 
the constraints for Einstein-aether theory are found.

\end{abstract}

\section{Introduction}

A field theory with diffeomorphism symmetry can not have deterministic field
equations. Although Einstein briefly thought this meant that a physical
theory could not be diffeomorphism invariant, he soon realized that 
the observables may be deterministically evolved, even though the fields are not\cite{Stachel}).
The solutions must involve four free functions  (in four spacetime dimensions), 
which correspond to
the freedom to apply an arbitrary time dependent diffeomorphism to the fields.
This means that out of the collection of Euler-Lagrange equations obtained from 
varying the fields in the action, there must be four combinations that 
are not independent of the others. 
These are initial value constraint
equations, and the remaining field equations imply that if the constraints hold initially,
they are automatically preserved in time.

This is all very familiar to anyone who has worked with general relativity,
but there is a twist if the matter 
energy-momentum tensor involves second time derivatives of the ``matter" fields,
as happens if the matter action involves covariant derivatives of tensor fields.
In that case, the initial value constraints are not just the $0\mu$ components
of the metric field equations;  instead, one must add to those a certain combination 
of the matter field equations. 
An example is Einstein-aether theory, in which the metric is coupled to a
unit timelike vector field. 

In this article the form of the constraint equations
in the presence of tensor matter is derived from a generalization 
of the Bianchi identity that includes the matter fields.
It is then applied to three examples
that illustrate different features. In particular, the constraints
for Einstein-aether theory are obtained. 
After writing this up I learned that the result is (of course) not new.
A paper by Bergmann\cite{Bergmann} from 1949 obtains the same result,
and it probably goes back even further.
It was recently derived by Seifert and Wald\cite{Seifert:2006kv},
and applied to Einstein-aether theory in \cite{Seifert:2007fr}.
However, it appears not to be common knowledge today.
Since theories with tensor matter are currently of interest,
it seems worth publishing this brief note, which aims to
explain the result and make it easily accessible using low brow,
coordinate based methods. 

\section{General relativity}

To begin with let us review how it works in
general relativity.

\subsection{Vacuum}
The vacuum Einstein equation,
\beq\label{efe}
G^{\a\b}=0,
\eeq
involves 10 independent components, matching 
the 10 independent components
of the spacetime metric. However 
4 of the 10 equations packaged into \eqref{efe} must not
actually be evolution equations. In fact, for any coordinate
system $x^\a=(x^0,x^i)$ ($i=1,2,3$), the $0\b$ components of 
\er{efe} involve no second derivatives with respect to $x^0$. 
(I will refer to $x^0$ as the ``time" in the following, although
in fact it can be any coordinate.) 
A simple way to see this\footnote{The argument 
is given this way
in Weinberg's 1972 book\cite{Weinberg}. 
It also appears in the 1975 english edition of 
Landau and Lifshitz's vol.\ 2\cite{LL},
which is translated from the 1973 russian 6th edition,
but not in the 1971 english edition. For a chronicle of 
the early history see Stachel\cite{Cauchy}.}
 is from the Bianchi identity,
\beq\label{Bianchi}
\nabla_\a G^{\a\b}=0,
\eeq
which in component form reads
\beq\label{Bianchicomp}
\p_0 G^{0\b} + \p_i G^{i\b}+ \G^\a{}_{\a\s}G^{\s\b} + \G^\b{}_{\a\s}G^{\a\s}=0. 
\eeq
If $G^{0\b}$ were to have second time derivatives, then the 
first term in \er{Bianchicomp} would have third time derivatives, unlike any other
term. This would be inconsistent with \er{Bianchicomp} being true as an identity.
Therefore we can infer that $G^{0\b}$ has no second time derivatives.
The equations 
\beq\label{Econ}
G^{0\b}=0
\eeq
are therefore
not evolution equations but rather constraints on initial data.
Moreover, the Bianchi identity is a first order differential
equation $\p_0 G^{0\b}=\dots$, which implies that
if these constraints hold at some initial time, and the 
rest of the field equations $G^{ij}=0$ hold at all times, then  
the constraints hold for all times.

It is no accident that the Bianchi identity is so intimately tied to the existence and preservation
of the constraint equations. That identity can be viewed as a direct consequence of 
the diffeomorphism invariance of the Einstein-Hilbert action, 
$S=\int R$. To reduce clutter, the metric volume element $\sqrt{-g}\, d^4x$ is
implicit here and in the following. 
(When the metric is varied, $\sqrt{-g}$ must certainly be varied as well.) 
Under a metric variation $\d g_{\a\b}$  the action varies as\footnote{We use the standard
notation: indices are lowered and raised with the metric and inverse metric, 
$R=g^{\a\b}R_{\a\b}$ is the Ricci scalar, $R_{\a\b}=R^\s{}_{\a\s\b}$ is the Ricci
tensor, $R^\s{}_{\a\b\g}$ is the Riemann tensor, and $G_{\a\b}=R_{\a\b}-\textstyle{\frac12}Rg_{\a\b}$ is the Einstein tensor.} 
\beq\label{dSg}
\d S =  -\int G^{\a\b}\, \d g_{\a\b} 
\eeq
If the metric variation arises from an infinitesimal diffeomorphism generated
by a vector field $\xi^\b$ we have 
\beq\label{dg}
\d_\xi g_{\a\b}={\cal L}_\xi g_{\a\b}=2\nab_{(\a}\xi_{\b)}
\eeq
where ${\cal L}$ denotes the Lie derivative, and the round brackets
on indices denote symmetrization. Substituting \er{dg} in \er{dSg},
integrating by parts, and assuming $\xi^\b$ vanishes at the boundaries,
we find that a diffeomorphism generates the
variation
\beq
\d_\xi S = \int (2\nab_\a G^{\a\b})\xi_{\b}.
\eeq
Since the Einstein-Hilbert action is a spacetime scalar,
$\d_\xi S$ must vanish for all $\xi^\b$, which implies the Bianchi identity
(\ref{Bianchi}). That is, the action is automatically stationary under 
those variations of the metric that arise from diffeomorphisms,
so the corresponding Euler-Lagrange equations do not impose 
any conditions. The identity implied by 
this symmetry indicates which four of the field equations
are constraints rather than evolution equations, and ensures that
those constraints are preserved in time.

\subsection{Matter}
Now let us consider what happens when matter is coupled to the metric.
The field equations then take the form 
\beq\label{efeT}
G^{\a\b}=8\pi T^{\a\b}
\eeq
(in units with $c=G=1$), where $T^{\a\b}$ is the matter stress tensor.
If the stress tensor involves no higher than first derivatives and the
matter field equations are of second order in derivatives, then the 
equations $G^{0\b}=8\pi T^{0\b}$ are again initial value constraint
equations. However, if the matter action involves the covariant
derivative, such as for example in a term like $(\nab_\a u_\b)(\nab^\a u^\b)$
(where $u^\b$ is a vector field), then variation of the metric
will produce (among other things) terms involving the second derivatives of 
the metric and matter fields, so $T^{\a\b}$ will generally
involve second time derivatives. Then 
$G^{0\b}=8\pi T^{0\b}$ will not be constraint equations. 

The difference from the vacuum case might seem puzzling since, 
when the matter satisfies its equation of motion, its 
stress tensor is divergence-free, $\nab_\a T^{\a\b}=0$. 
Together with the Bianchi identity this yields the
equation $\nab_a(G^{\a\b}-8\pi T^{\a\b})=0$, and it is tempting to
argue by analogy with the Bianchi identity case 
that validity of this equation implies that 
$G^{0\b}-8\pi T^{0\b}$ can have no second time derivatives.
This would be incorrect however, since the equation uses 
the equations of motion of the matter field, so does not 
hold {\it identically} for all matter fields. In order to discover the
true constraint equations in theories like this, what we need to do is
find the analog of the Bianchi identity, that is, the identity that 
follows from the diffeomorphism invariance of the action.

\section{Any metric theory with tensor matter}
Consider then a diffeomorphism invariant action $S[g,\psi]$ that
is a functional of the metric $g_{\a\b}$ and some tensor matter field
of any rank, $\psi_{\a\dots}^{\b\dots}$. 
For notational simplicity I will assume $\psi$ is a $(1,1)$ tensor
in what follows but it should be obvious how to modify the equations
to allow for the general case.
A diffeomorphism induces the
variation (\ref{dg}) for the metric and 
\beq\label{dpsi}
\d_\xi \psi_{\a}^{\b}={\cal L}_\xi \psi^{\a}_{\b}
=\xi^\g\, \nab_\g\psi^{\a}_{\b} +\psi_{\g}^\b\,  \nab_\a\xi^\g{}
-\psi_\a^\g \, \nab_\g
\xi^\b{}_{}
\eeq
for the tensor. 
Actually the Lie derivative is independent of the metric, so all the
Christoffel symbols in (\ref{dpsi}) cancel out, but to maintain manifest covariance
and to avoid the need to explicitly write the $\sqrt{-g}$ factor it is more convenient 
to use covariant derivatives.

The variation of the action under a diffeomorphism is
\beq\label{dS}
\d_\xi S = \int E^{\a\b}\, \d_\xi g_{\a\b} + H^\a_\b\,  \d_\xi \psi_{\a}^{\b},
\eeq
where 
\beq
E^{\a\b}=\frac{\d S}{\d g_{\a\b}},\qquad H^\a_\b=\frac{\d S}{\d\psi_{\a}^{\b}}.
\eeq
$E^{\a\b}=0$ is the metric field equation, 
and $H^\a_\b=0$ is the matter field equation. Now we insert the
diffeo variations (\ref{dg}) and (\ref{dpsi}) into (\ref{dS}), integrate by
parts on all derivatives of $\xi^\a$, and redefine some dummy indices
to obtain
\beq
\d S = \int \left[\nab_\a(-2E^\a_\b -H^\a_\g\psi^\g_\b 
+ H^\g_\b\psi^\a_\g)-H^\a_\g \nab_\b\psi^\g_{\a}\right] \xi^\b.
\eeq
Since the action is diffeomorphism invariant, this variation
must vanish for all vector fields $\xi^\b$, so we infer an 
identity,
\beq\label{identity}
\nab_\a(-2E^\a_\b -H^\a_\g\psi^\g_\b 
+ H^\g_\b\psi^\a_\g)=H^\a_\g \, \nab_\b\psi^\g_{\a}
\eeq
This generalizes the Bianchi identity (\ref{Bianchi}). Note that, 
unlike the latter, it 
is not just the statement that a certain tensor is divergence-free,
but rather that the divergence is equal to some other quantity. 
It includes the matter fields and holds off-shell, i.e.\ for all field
configurations.  

Now to identify the initial value constraint equations, we
expand the covariant divergence on the left hand side of 
(\ref{identity}) in terms of partial derivatives and Christoffel
symbols as in (\ref{Bianchicomp}). The term 
$\del_0(-2E^0_\b -H^0_\g\psi^\g_\b + H^\g_\b\psi^0_\g)$
will have one higher time derivative order than any other 
term in \er{identity}.
Since \eqref{identity} holds identically for all fields, we infer 
that the quantity 
\beq\label{0comp}
-2E^0_\b -H^0_\g\psi^\g_\b + H^\g_\b\psi^0_\g
\eeq
has time derivative order that is one less than the highest 
order in the field equations $E^{\a\b}=0$ and $H^\a_\b=0$.
Also, it is a combination of components of the field equations,
so it vanishes when the field equations hold. We thus arrive 
at the conclusion that the initial value 
constraints are given by 
\beq\label{result}
{\cal C} _\b\equiv -2E^0_\b -H^0_\g\psi^\g_\b + H^\g_\b\psi^0_\g=0.
\eeq 
For a general rank tensor
matter field $\psi^{\cdots}_{\cdots}$ 
there would be one term in $\cal C_\b$ for each index. 
Each covariant matter index contributes 
$-H^{\cdots 0\cdots}_{\cdots}\psi^{\cdots}_{\cdots\b\cdots}$,
and each contravariant index contributes 
$H_{\cdots \b \cdots}^{\cdots}\psi_{\cdots}^{\cdots0\cdots}$.
For multiple matter fields, there would be similar terms
for each matter field. 
The generalized Bianchi
identity (\ref{identity}) implies that the constraints 
are automatically preserved in time if they
hold at one time and the 
matter field equations and other components
of the metric field equations hold at all times.

\section{Examples}
We now apply this result to three examples.
The first example is Einstein-Maxwell theory.
The matter field is a covariant vector potential $A_\a$, 
and the matter Lagrangian is $-\textstyle{\frac14}F^{\a\b}F_{\a\b}$,
with $F_{\a\b}=\del_\a A_{\b}-\del_\b A_{\a}$.
The Maxwell stress tensor
has no second derivative terms, so 
the quantities $E^{0\b}=-G^{0\b}+8\pi T^{0\b}$ involve no
second time derivatives. Nevertheless, it is also true that
$C_\b$ in (\ref{result}) has no second time derivatives.
The extra term in $C_\b$ in this case is 
$-H^0 A_\b$, so it must be that also 
$H^0$ has no second 
time derivatives. Indeed, 
$H^0=\nab_\a F^{\a0}{}_{}$, and 
the antisymmetry of $F^{\a\b}$ implies that 
the $\a$ index must be spatial.
This field equation  corresponds to 
$\nabla\cdot{\bf E}=0$, the initial value
constraint for Maxwell's equations.

For the case of Einstein-aether theory\cite{Jacobson:2008aj}
 the tensor matter
is the aether vector $u^\a$, and the Lagrangian 
contains Christoffel symbols, for instance in the term  
$(\nab_\a u_{\b})(\nab^\a u^{\b})$, so the stress tensor has second time
derivatives. The constraint equations are
\beq\label{aether}
{\cal C} ^{\rm \ae ther}_\b\equiv -2E^0_\b  + u^0H_\b=0,
\eeq 
where $H_\b=0$ is the aether field equation, which
contains second time derivatives that cancel those in 
$E^0_\b$ when the two are added with the coefficients
prescribed in \er{aether}. 
A special feature of this theory is that 
the field equation has no component in the direction
of $u^\b$, i.e.\ $H_\b u^\b=0$. That is because 
$u^\b$ is constrained to be a unit timelike vector, 
hence all of its variations are
orthogonal to it, $u_\b \d u^\b=0$.\footnote{If the unit
constraint is imposed by variation of a Lagrange multiplier, this
statement holds only after the multiplier is solved for and 
inserted back into $H_\b$.} Thus the $0$-$u$ component
of the Einstein equation, $E^0_\b u^\b=0$, is an initial value constraint 
without any additional terms involving the aether field equation.
In particular, on a hypersurface to which the aether is orthogonal,
the $uu$ component of the aether stress tensor has no second derivatives in the 
$u$ direction. 
 
The final example is Ho\v{r}ava gravity\cite{Horava:2009uw,Sotiriou:2010wn}.
I will first describe it
in the original $3+1$ formulation with a preferred time, 
and then explain how the constraints work in its fully covariant formulation. 
In the original form, 
this theory has a fixed time function, defined up to reparameterizations,
that determines a foliation of spacetime. The remaining symmetry is 
the 
time dependent spatial diffeomorphisms on the 
constant time surfaces. 
The field variables are the spatial metric, the shift vector, and the 
lapse scalar.
I will discuss only
the version of this theory in which the lapse $N$ is allowed to depend
on the spatial coordinates, and in which the action contains
all terms up to two derivatives, including\cite{Blas:2009qj} a term 
proportional to $(\del_i N)(\del^i N)/N^2$ 
(which is invariant under time dependent rescaling of $N$
since $\del_i$ denotes the spatial gradient). 
This theory has three first class
constraints that generate spatial diffeomorphisms in the 
Hamiltonian formulation, and a fourth constraint that is
second class, because the theory
lacks surface deformation symmetry except for global time 
reparameterization\cite{Kluson:2010nf,
Donnelly:2011df,Bellorin:2011ff}.\footnote{On a 
compact spatial manifold there
is one global first class surface deformation constraint
that generates time reparameterizations\cite{Donnelly:2011df}.}
This fourth constraint is the field equation for the lapse,
and is an elliptic equation in spatial derivatives,
so determines the lapse in terms of the other variables.

In the covariant formulation\cite{Blas:2009qj,Jacobson:2010mx}, the field variables are the spacetime metric
and a scalar field $T$, with a symmetry under reparameterizations of 
$T\rightarrow f(T)$. The two derivative action is just that of Einstein-aether
theory, with aether vector $u_\a=N \del_\a T$ and 
$N=(g^{\a\b}\del_\a T\, \del_\b T)^{-1/2}$. 
Because the ``matter" field $T$ is a scalar, the generalized Bianchi
identity (\ref{identity}) takes the form
\beq\label{Hidentity}
\nab_\a(-2 E^\a_{\b})= H\, \nab_\b T,
\eeq
where $H=-\nabla_\a[N(\d^\a_\b-u^\a u_\b)\d S/\d u_\b]$ 
is the $T$ field equation\cite{Jacobson:2010mx}.
The covariant derivative
of $u^\a$ appears in the action so, although the matter is a scalar field,
the aether part of 
$E^\a_\b$ has derivatives of order two on $g_{\a\b}$ and three
on $T$, while $H$ has derivatives of one higher order on each
field variable. It is not until the gauge $x^0=T$ is chosen that the time 
derivative order of the theory drops to two:
the field variable $T$ disappears
from the equations, and the derivative orders of $E^\a_\b$ and $H$
remain two and three respectively, but the third derivative in the $H$ field equation is always in a spatial direction. Since both field
equations are now second order in time derivatives, 
the gauge-fixed identity (\ref{Hidentity}) implies that all four
quantities $E^0_\b$ have
no second time derivatives. 
(The right hand side of (\ref{Hidentity}) 
has no component orthogonal to $u^\a$ anyway, so even 
if $H$ had third time derivatives, one could have still inferred that
the time--space components $E^0_i$ have no second time derivatives.)
 That is, the constraints are of the usual
form, with no need to add a term involving the matter field equation.
 
\section{Discussion} 

The constraint equations are essential input for  
calculation of evolution from initial data.
That evolution could be in a timelike direction, as in ordinary time
evolution, or a spacelike direction,
as e.g.\ in the case of finding static or stationary solutions to field equations.
In particular, the constraint equations for Einstein-aether theory 
on a timelike surface of constant radial coordinate were used in \cite{Barausse:2011pu}
to find static, spherically symmetric black hole solutions to that theory.
The constraints on a spacelike hypersurface
in spherical symmetry were used  in 
\cite{Seifert:2007fr} to study perturbative stability 
of solutions in various modified gravity theories including Einstein-aether theory.
The constraints for 
a spherically symmetric, spacelike hypersurface to which 
the aether is orthogonal were worked out for Einstein-aether theory in 
\cite{Garfinkle:2007bk}, and used in a numerical simulation of 
collapse of  matter to form a black hole. An interesting
application of the generalization beyond spherical symmetry 
would be to numerical simulation of axisymmetric black hole collapse, which 
could yield at late times a stationary rotating black hole solution.

\section*{Acknowledgments}
I thank E.~Barausse and T.~Sotiriou for helpful discussions, and D.~Garfinkle for useful comments on a draft of this article.
This work was supported in part by the National Science Foundation
under grants  PHY-0601800 and PHY-0903572.

\end{document}